\documentclass[prd,preprintnumbers,nofootinbib,aps,9pt]{revtex4}
\usepackage{subeqnarray}
\usepackage{epsfig,amsmath,amssymb}
\usepackage{mathrsfs}
\usepackage[usenames,dvipsnames]{color}
\usepackage[pagebackref=true, colorlinks=true]{hyperref}
\definecolor{redish}{rgb}{0.7,0.2,0.0}  
\definecolor{bluish}{rgb}{0.2,0.5,0.8}
\hypersetup{linkcolor=redish,          
                  citecolor=blue,        
                  filecolor=magenta,      
                  urlcolor=bluish}          
\numberwithin{equation}{section}

\newcommand{\ba}{\nopagebreak[3]\begin{eqnarray}}
\newcommand{\ea}{\end{eqnarray}}
\newcommand{\nn}{\nonumber}
\newcommand{\f}{\frac}
\def \p{\partial}
\def \b{\beta}

\def \g{\gamma}
\def \d{\delta}

\def \lp{\ell_p}
\def \j{\sqrt{j(j+1)}}

\def \lm{\lambda}

\def \s{\sigma}

\def \Q{N_0}
\def \H{\mathcal{H}}
\def \P{\mathcal{P}}
\def \sj{s_j^{\star}}
\begin{document}
\title{Thermodynamic partition function from quantum theory for black hole horizons in loop quantum gravity} 
\author{Abhishek Majhi}%
\email{abhishek.majhi@gmail.com}
\affiliation{School of Physics, Indian Institute of Science Education and Research\\ Thiruvananthapuram (IISER TVM), Trivandrum 695016, IISER Trivandrum, India\\}
\affiliation{Astro-Particle Physics and Cosmology Division\\Saha Institute of Nuclear Physics\\Kolkata, India}
\begin{abstract}
We establish the link between the thermodynamics and the quantum theory of black hole horizons through the construction of the thermodynamic partition function, partly based on some physically plausible arguments, by beginning from the description of  quantum states of the horizon, considering loop quantum gravity(LQG) as the underlying theory. Although, the effective `thermalized\rq{}  form of the partition function has been previously used in the literature to study the effect of thermal fluctuations of the black hole horizon, nonetheless the direct link to any existing quantum theory (which is here taken to be LQG), especially a derivation of the partition function from the quantum states of the horizon, appears to be hitherto absent.
This work is an attempt to bridge this small, but essential, gap that appears to be present between the existing literature of quantum theory and thermodynamics of black holes. Further, it may be emphasized that this work is {\it only} concerned with the {\it metric independent} approaches to black hole thermodynamics.

\end{abstract}
\maketitle
\section{Introduction}\label{sec1}
The thermodynamic properties associated with spacetime geometry are intimately associated with the presence of a physical horizon. In the Euclidean quantum gravity approach, it had been actually proven that the entropy of the spacetime geometry vanishes in the absence of a horizon\cite{hawgib}. Similarly, using certain properties of Canonical quantum gravity, within certain approximations, it has been shown that the partition function of a spacetime reduces to that of the horizon only\cite{th1,th2}. The physical significance of the presence of the horizon looks even more compelling when we find that i) the entropy of the black hole is given by it horizon area instead of the volume of the region behind the horizon\cite{bhal} ii) the thermal radiation of the black hole is emitted at the temperature which is the one associated with the horizon\cite{hrad,wrad} iii) there are laws of thermodynamics associated with the horizon\cite{laws}. All these findings compel us to think that there may be some quantum theory of the horizon geometry itself which describes the independent quantum degrees of freedom of the horizon and then only we can have a proper understanding and logical reasoning behind the emergent thermodynamic properties of the horizon manifested in its own right. The knowledge of such quantum theory of the horizon will also enable us to formulate a spacetime-metric or back ground independent horizon thermodynamics. In fact, in recent literature there has been some serious attempts to study horizon thermodynamics without referring to the bulk spacetime by constructing the partition function for the horizon only and investigating the relevant thermodynamic properties.

Thermal fluctuations of black hole horizons have been studied in various details by several authors in the recent past leading to interesting results regarding black hole  thermodynamics\cite{dmb,cm4,gomed,cm1,th1,cm2,th2,cm3}. In \cite{th1,th2,cm3} thermal stability of black holes under thermal fluctuations were studied in canonical and grand canonical ensembles leading to a stability criterion for black holes which successfully explained the stabilities and instabilities of certain known black hole solutions. On the other hand, the issues of canonical and grand canonical entropy of black holes in presence of thermal fluctuations have been studied in \cite{dmb,cm4,gomed,cm1,cm2}. The common and unique property of all these works is the use of the horizon partition function, whose construction is based on some heuristic modeling or assumptions or approximations regarding the underlying quantum  structure of the horizon geometry, without referring to any classical background geometric structures, such as the metric and hence, valid for arbitrary black hole horizons independent of the structure of the associated bulk spacetime. The thermodynamic properties associated with black hole horizons urge one to think of an inherent quantum description of the horizon geometry which will give rise to the notion of microstates so as to have a physical meaning of the associated entropy and hence, thermality. This is the underlying physical essence of this particular statistical mechanical approach to horizon thermodynamics presented and investigated in \cite{dmb,cm4,gomed,cm1,th1,cm2,th2,cm3}. 

However, none of these works\cite{dmb,cm4,gomed,cm1,th1,cm2,th2,cm3} presents a logical straightforward derivation of the partition function based on the fundamental quantum structure of black hole horizons, which would have made the link between the microscopic quantum theory of the horizon and the associated thermodynamics as had been urged by the works in \cite{hawgib,hrad,wrad,bhal,laws}. This crucial direct link still goes begging, even though there is a well defined quantum theory for the horizon geometry  present in the literature for several years\cite{qg1,qg2}. Perhaps the clearest quantum description of the geometry of a black hole horizon is provided by loop quantum gravity\cite{qg1,qg2}, where the  
quantum degrees of freedom of the horizon belong to the Hilbert space of a quantum Chern-Simons theory coupled to point like sources (to be explained in detalis later).

The theory, thus, provides a self-contained platform for the application of the statistical mechanical techniques so as to unravel the corresponding thermodynamic properties of the black hole horizon. This paper is aimed at providing a direct link between the quantum theory of horizon geometry and the associated thermodynamics through  a derivation of the thermodynamic partition function associated with the horizon. This bridges the gap between the quantum geometric framework of equilibrium horizons laid down in \cite{qg1,qg2} and the thermodynamic aspects of the horizon studied in \cite{dmb,cm4,gomed,cm1,th1,cm2,th2,cm3}, which in turn, answers the issues motivated by the works in \cite{hawgib,hrad,wrad,bhal,laws} as discussed earlier. This is precisely the important value addition of this work to the extant literature and hopefully may be of interest to the concerned readers. The structure of the paper can be debriefed as follows.

\section{Equilibrium horizon and quantum fluctuations : microcanonical ensemble}
\label{sec2}
In modern day literature, a black hole horizon in gravitational and thermal equilibrium with ambient matter and radiation is described by the notion of an isolated horizon (IH)\cite{ih1,ih2,ih3}, which is a  2+1 dimensional null inner boundary of a 3+1 dimensional spacetime. An IH is a generalization of an event horizon of a stationary black hole spacetime to a more realistic scenario. As opposed to the global notion of the event horizon, an IH is defined locally, without any reference of the ambient bulk spacetime and hence, can admit matter and radiation arbitrarily close to itself, whereas an event horizon can not. The laws of black hole mechanics are completely realizable in the local framework of IH\cite{ih3,ih4,ih5}. The mass and the surface gravity associated with the IH are also defined from a completely local perspective\cite{ih3}. The strength of the IH framework is its local, background independent description in terms of the connection variables, which has led to rigorous analysis of the connection dynamics, symplectic structure, etc. of the IH. The analysis reveals that the symplectic structure of the IH is that of an $SU(2)$ {Chern-Simons}({CS}) theory\cite{ih1,ih2}\footnote{A concerned reader may look into \cite{km11} where, taking the event horizon in the Schwarzschild spacetime as an example(which is a special case of an IH), it has been shown that there is indeed an $SU(2)$ CS theory on the event horizon.}. In the quantum theory, the states of a  QIH  are given by that of the CS theory coupled to the edges of the bulk spin network which span the bulk quantum geometry and intersect the IH at specific points called punctures\cite{qg1,qg2}.

The Hilbert space of a quantum spacetime admitting QIH as an inner boundary is given by ${\cal H}={\cal H}_V\otimes{\cal H}_S$ modulo gauge transformations, where $V$ denotes bulk and $S$ denotes boundary(QIH) at a particular time slice\cite{qg1,qg2}. Mathematically, if the 4d spacetime $({\bf R}\times\Sigma)$ admits a 3d IH $(\Delta)$ as null inner boundary, then $S\equiv \Delta\cap\Sigma$ denotes a cross-section of the IH \cite{ih1,ih2}.  Hence, a generic quantum state of the spatial geometry of such a spacetime can be written as $|\Psi\rangle=|\Psi_V\rangle\otimes|\Psi_S\rangle$, where $|\Psi_V\rangle$ is the wave function corresponding to the volume$(V)$ or bulk states represented by an oriented graph, say $\Gamma$, consisting of edges and vertices \cite{ashlew} and $|\Psi_S\rangle$ denotes a generic quantum state of the QIH. $|\Psi_S\rangle \in{\cal H}_S\equiv$ the Hilbert space of the CS theory coupled to the punctures $\{\P\}$ made by the bulk spin network $\Gamma$ with the IH endowing them with the spin representations carried by the respective piercing edges which are solely responsible for all the relevant features of the QIH, the most important being the quantum area spectrum of the QIH. To be precise, for a given $N$ number of punctures, with spins $(j_1,\cdots,j_N)$, the QIH Hilbert space is given by $\H_S\equiv \text{Inv}(\otimes_{i=1}^N\H_{j_i})$ where `Inv' denotes the invariance under the local $SU(2)$ gauge transformations on the QIH. Now, as it is seen that at the quantum level the full Hilbert space is the direct product space of the bulk and boundary Hilbert spaces, a generic quantum state of the QIH (boundary) can be written in terms of basis states on $\H_S$, independent of the bulk wave function. Hence, one should understand that a basis state of the QIH Hilbert space is actually a generic quantum state of the full Hilbert space, since the bulk part of the wave function is a linear combination of the basis states of the bulk geometry. In other words, a given {\it spin configuration} on the QIH admit all possible graphs $(\Gamma$-s) in the bulk consistent with the given configuration. This spin configurations provide the area eigenstate basis, which is the all important material in the context of QIH entropy. Such a basis state of the QIH Hilbert space is denoted by the ket $|\{s_j\}\rangle$. This is an eigenstate of the area operator associated with the QIH, having the area eigenvalue given by $\hat A_S|\{s_j\}\rangle=8\pi\g\lp^2\sum_{j=1/2}^{k/2} s_j\j|\{s_j\}\rangle$. Such a spin configuration (eigenstate) has a $(N!/\prod_j s_j!)$-fold degeneracy due to the possible arrangement of the spins yielding the same area eigenvalue. Hence, a generic quantum state of the QIH can be written as 
\ba
|\Psi_S\rangle=\sum_{\left\{s_j\right\}}c[\left\{s_j\right\}]|\left\{s_j\right\}\rangle\nn
\ea 
where $|c[\left\{s_j\right\}]|^2=\omega[\left\{s_j\right\}]$(say) is the probability that the QIH is found in the state $|\left\{s_j\right\}\rangle$.

We would like to mention that the topological structures i.e. the punctures, arise only in the quantum theory and we can consider it to be a macroscopic parameter only if we deal with an ensemble of QIHs rather than an ensemble of IHs. Thus, the Hilbert space of a QIH for a given $N$ should be appropriately written with proper designation as $\H_{S}^{k,N}=\text{Inv} \left(\bigotimes_{l=1}^N \H_{j_l}\right)$. But, the full Hilbert space of a classical IH designated by the corresponding CS level $k$  takes into account all possible sets of punctures\cite{qg1,qg2} and  is given by
\ba
\H_{S}^{k}&=&\bigoplus_{\{\P\}}\text{Inv} \left(\bigotimes_{l=1}^N \H_{j_l}\right)
~~~~\text{where} ~~~ \bigoplus_{\{\P\}}\equiv \bigoplus_{N ; \frac{1}{2}\leq j_l\leq\f{k}{2}\forall l\in[1,N] \ni \sum_{l=1}^N\sqrt{j_l(j_l+1)}=\f{k}{2}\pm\mathcal{O}(\f{1}{8\pi\g})}\nn
\ea
and $\g$ is the Barbero-Immirzi parameter. The CS level$(k)$ is defined as  $k\equiv A/4\pi\g\lp^2$  where $A$ is the classical area of the IH \cite{qg1}. In the quantum theory, the CS coupling $k$ is an integer which is a necessary pre-quantization condition for the quantization of the classical IH\cite{qg2}. Since, we shall begin with the application of quantum mechanical ensemble theory
to the QIH, the macroscopic parameters will be considered to be $k$ and $N$\cite{ampm2} and the microstates resulting in the quantum degeneracy for a given $k$ and $N$ will give rise to the statistical or microcanonical entropy as a function of $k$ and $N$\cite{ampm2,meih}. 

\section{Quantum Fluctuations : The Microcanonical Ensemble }
\label{sec3}
The number of microstates for a QIH with CS level $k$ and number of punctures $N$ is given by the dimension of the Hilbert space $\H_{S}^{k,N}=\text{Inv} \left(\bigotimes_{l=1}^N \H_{j_l}\right)$ where $\frac{1}{2}\leq j_l\leq\f{k}{2}\forall l\in[1,N]$.  The formula for the number of microstates for an arbitrary  sequence of spins $(j_1, ...., j_N)$ on $N$ number of punctures on the QIH is given by \cite{km98} 
\ba
\Omega(j_1, \cdots ,j_N)=\f{2}{k+2}\sum^{k+1}_{a=1}\f{\sin\f{a\pi(2j_1+1)}{k+2}\cdots\sin\f{a\pi(2j_N+1)}{k+2}}{\left(\sin\f{a\pi}{k+2}\right)^{N-2}}\label{cs}
\ea
To obtain the total number of microstates of a QIH for a given $k$ and $N$ i.e. the dimensionality of $\H^{k,N}_S$, we must consider all possible values of spin from $1/2$ to $k/2$ for each puncture. This is the precise argument which was provided in \cite{ampm2} in the other way round so as to render $k$ and $N$ to be macroscopic parameters designating a macroscopic state of a QIH i.e. taking the sum over all spins, the only parameters which are left bare are $k$ and $N$. Hence we can write
$\Omega(k,N)=\sum_{j_1,\cdots ,j_N}~\Omega(j_1, \cdots ,j_N)$ where each spin is summed from $1/2$ to $k/2$. In principle and by definition, the microcanonical entropy of the QIH is given by $S_{MC}=\log \Omega(k,N)$, where we have set the Boltzmann constant to unity. In practice\cite{ampm2}, the calculation of $\Omega(k,N)$ is performed only approximately\footnote{The approximation is good enough in the limit $k,N\to\infty$ which is the appropriate for QIHs with large classical area and large number of punctures.} by finding the {\it most probable distribution} through the application of the method of Lagrange undetermined multipliers. The calculation has been extensively carried out in \cite{ampm2} and is needless to repeat here.  What we shall discuss here is the physical essence of the method alongside mentioning the crucial steps of the calculation. Firstly, one can express the sum over spin values  as sum over spin configurations by the application of the multinomial theorem\cite{ampm2} i.e.
\ba
\Omega(k,N)=\sum_{j_1,\cdots ,j_N}~\Omega(j_1, \cdots ,j_N)=\sum_{\left\{s_j\right\}}~\Omega[\left\{s_j\right\}]
\ea
 where 
\ba
\Omega[\left\{s_j\right\}]=\f{N!}{\prod_{j}s_j!}\f{2}{k+2}\sum^{k+1}_{a=1}\sin^2\f{a\pi}{k+2}\prod_{j}\left\{\f{\sin\f{a\pi(2j+1)}{k+2}}{\sin\f{a\pi}{k+2}}\right\}^{s_j}
\ea
This puts forward a very clear picture of the physical scenario which manifest the quantum jumps of QIH state from one area eigenstate $|\{s_j\}\rangle$ to another, since the spin configurations are the eigenstates of the area operator of the QIH as discussed earlier. The variation of spin values all over the punctures of the QIH are the quantum fluctuations of the system. Now, any puncture can take any spin value from $1/2$ to $k/2$ with equal a priori probability and hence, all the spin sequences are equally probable. But when we look at the spin configurations, the one allowing the maximum number of possible equi-probable  microstates(spin sequences) i.e. maximally degenerate,  is the most probable spin configuration. Hence quantum state corresponding to the most probable configuration is the highest entropy state. At par with the basic postulates of equilibrium statistical mechanics, the degeneracy corresponding to this state overwhelmingly outnumbers the degeneracies corresponding to the other subdominant configurations which may be regarded as negligible quantum fluctuations. Thus, it is a good enough approximation to consider $\Omega(N,k)\simeq\Omega[\{s_j^{\star}\}]$, where $s_j^{\star}$ is the distribution corresponding to the most probable spin configuration and comes out to be
\ba
s_j^{\star}\simeq N (2j+1)\exp[-\lm\j -\s]\label{mpd2}
\ea
It may be mentioned that the classical area $(A_{cl})$ is related to $k$ as $k=A/4\pi\g\lp^2$. It is straightforward to show that the microcanonical entropy comes out to be  \cite{meih,ampm2}
\ba
S_{MC}=\lm k/2+N\sigma
\ea
where $\lm$ and $\s$ are the Lagrange multipliers and we have neglected  the subleading terms in the entropy\cite{ampm2} that are irrelevant in the present analysis. The Lagrange multipliers are solvable in terms of $k$ and $N$ from the following equations\cite{meih,ampm2}
\begin{subeqnarray}\label{apcons}
e^{\s}&=&\f{2}{\lm^2}\left(1+\f{\sqrt 3}{2}\lm\right)e^{-\f{\sqrt 3}{~2}\lm}\label{siglm}\\
\f{k}{N}&=&1+\f{2}{\lm}+\f{4}{\lm(\sqrt 3\lm+2)}\label{lmro}
\end{subeqnarray}
It should be noted that the most probable spin configuration is determined by the given set of values of $k$ and $N$.

The physical essence of the microcanonical ensemble analysis is that the thermal fluctuations of the macroscopic variables are completely disallowed and what we deal with are purely quantum mechanical fluctuations of the system. It is easy to understand because we define the microcanonical ensemble by fixing the macroscopic variables, which are here $k$ and $N$, and do not allow them to fluctuate. Thus the classically behaving macroscopic variables remain constant and the QIH jumps  randomly from one quantum mechanical state to another which gives rise to the quantum uncertainty and hence, the microcanonical entropy. In that sense, the resulting entropy is purely statistical. As we shall pass on to the grand canonical ensemble scenario, the effects of the thermal fluctuations will be taken into account because now the fixed macroscopic variables will be allowed to fluctuate.   

{\bf Note :} At this point, having known the degeneracy corresponding to given $k$ and $N$, we can write down the canonical partition function as
\ba
Z_C(\b,N)~=~\sum_{k} \Omega(k,N)\exp \left[-\b E(k,N)\right] \label{tezc}
\ea
and the  grand canonical partition function as 
\ba
Z_G(\b,\mu)~=~\sum_{k,N} \Omega(k,N)\exp -\b\left[E(k,N)-\mu N\right]\label{tezg}
\ea 
where $E(k,N)$ is the mean energy or the statistical mean of the Hamiltonian operator for the QIH\footnote{Although the classical energy associated with the classical IH \cite{ih3} has  not been quantized till date, there has been a proposal of the most general possible structure of the Hamiltonian operator associated with the QIH\cite{me6,me7}, which satisfies the properties of the classical energy in the correspondence limit.} with given $k$ and $N$. This is like considering several different copies of microcanonical ensembles of QIHs designated by different sets of values of $k$ and $N$ and then dealing with the effective classical behaviour of the relevant quantities, the quantum fluctuations being averaged out\cite{landau}. So eq.(\ref{tezc}) and eq.(\ref{tezg}) can be regarded as the `thermalized' forms of the quantum mechanical canonical and grand canonical partition functions respectively, since the many body structure of the QIH is not manifested at all in these above forms and are glossed out by the classically behaving macroscopic variables. {\it Similar} forms of the partition functions have been extensively used to investigate the  thermodynamics of black holes in the IH framework previously in literature \cite{dmb,gomed,cm1,cm2,cm3,th1,th2}. However, there are several technical caveats in common which plague the approaches which also curtains some important physical issues regarding the quantum statistics and thermodynamics of black holes. In the forthcoming sections, we shall arrive at the above `thermalized' forms of the partition functions by beginning from the treatment with quantum mechanical ensembles of QIHs to get an in depth understanding of the thermodynamical issues based on the underlying quantum structures. Remarkably, albeit expectedly, the technical caveats of the earlier approaches to black hole thermodynamics get removed automatically, alongside the fact that the associated physical understandings become far more transparent, which will be discussed in details in a separate section.

\section{Thermalization : Other Quantum Mechanical Ensembles}
\label{sec4}
Having reviewed the microcanonical ensemble results for a QIH, now we shall investigate the other quantum mechanical ensembles viz. canonical and grand canonical. In the following paragraphs we shall explore in details the construction of the canonical and grand canonical partition functions beginning from the very basic definitions and arrive at their respective `thermalized' forms.

\subsection{Quantum Mechanical Canonical Partition Function} 
\label{subsec4.1}
The Hilbert space structure $\H_S^{k,N}$ of a QIH with given $k$ and $N$ emulates that of a gas of particles\cite{GP} in the sense that every physical phenomenon concerning a QIH is a collective many body effect, a puncture being an individual body. It is also evident from the quantum area of QIH i.e. $8\pi\g\lp^2\sum_{l=1}^N\sqrt{j_l(j_l+1)}$\cite{qg1,qg2} or the quantum energy of the QIH proposed in \cite{me6,me7}.  Following this quantum `particle-like' structure of the QIH, the quantum mechanical canonical partition function can be written as 
\ba
Z_C(\b,N)=\sum_{\{s_j\}}~\Omega[\{s_j\}]\exp[-\b E[\{s_j\}]]\label{zcqm}
\ea
where $E[\{s_j\}]=\sum_j\epsilon_js_j$ is the quantum energy of the QIH corresponding to the spin configuration $\{s_j\}$, $\epsilon_j$ is the energy contribution from a single puncture carrying spin $j$ and for any configuration,  the total number of punctures must add up to $N$ since it is kept fixed in the canonical ensemble. 

Now, from eq.(\ref{mpd2}), considering that $\lm$ and $\s$ are solvable in terms of $k$ and $N$ from eqs.(\ref{apcons}), it is manifest that there is an $\sj$ for every $k$ and $N$. Thus, considering a fixed value of $N$, a different choice of $k$ will give a different $\sj$. It is then completely physical to argue that the most probable distributions corresponding to $k-1$ and $k+1$ are the closest less probable distributions corresponding to the most probable distribution for $k$. Likewise, the other distributions in eq.(\ref{zcqm}) are the most probable distributions for some $k$. Thus, for fixed $N$, in the quantum mechanical canonical ensemble, the sum over spin configurations can be replaced by sum over $k$  i.e.
\ba
Z_C(\b,N)=\sum_{k}~\Omega(k,N)\exp[-\b E(k,N)]\label{zcth}
\ea
where $\b$ is the temperature of the heat bath with which the QIH is considered to be in thermal contact. Since $k$ is an integer and since we deal with the large $k$ regime only the above expression can  be well approximated by replacing with the summation by an integration as follows 
\ba
Z_C(\b,N)\simeq\int~dk~\Omega(k,N)\exp[-\b E(k,N)]\label{zcapp}
\ea
Eq.(\ref{zcapp}) is the expression for the canonical partition function for a QIH
which is suitable for studying the thermal fluctuations. This may be regarded as the `thermalized' form of the quantum mechanical canonical partition function given by eq.(\ref{zcqm}). The information about the many-body quantum structure of the QIH is glossed by the classical behaviour of the the macroscopic variables.

\subsection{Quantum Mechanical Grand Canonical Partition Function} 
\label{subsec4.2}
Now, we shall allow the macroscopic variable $N$ to vary along side the energy. So, let us consider a grand canonical ensemble of QIHs where the number of punctures $N$ is now allowed to vary along side the energy. The quantum mechanical grand canonical partition function can be written as 
\ba
Z_{G}(\b,\mu)=\sum_{N=1}^{\infty}\exp[\b\mu N]Z_C(\b,N)\label{zgqm}
\ea 
where $Z_C$ is the partition function for a quantum mechanical canonical ensemble of QIHs. $\mu$ is some fictitious parameter conjugate to the macroscopic variable $N$, playing a role {\it analogous} to the chemical potential in case of a gas of particles. Let us rewrite eq.(\ref{zgqm}) manifesting the many-particle structure of the QIH as 
\ba
Z_{G}(\b,\mu)&=&\sum_{\{s_j\}}~\Omega[\{s_j\}]\exp[-\b E[\{s_j\}]+\b\mu N[\{s_j\}]]
\ea 
Now, we can argue in the similar way as in case of the canonical ensemble. From eq.(\ref{mpd2}), considering that $\lm$ and $\s$ are solvable in terms of $k$ and $N$ from eqs.(\ref{apcons}), it is manifest that there is an $\sj$ for every $k$ and $N$. Thus,  different choices of $k$ and $N$ will give a different $\sj$. Based on a physical argument similar to that of the quantum mechanical canonical ensemble,  in the quantum mechanical grand canonical ensemble, the sum over spin configurations can be replaced by sum over $k$  and $N$ i.e.
\ba
Z_G(\b,\mu)~=~\sum_{k,N} \Omega(k,N)\exp -\b\left[E(k,N)-\mu N\right] 
\ea 
$k,N$  are like `effective' quantum numbers for the energy and  they are equispaced as they are integers.  Since we are working in the large $k$ and large $N$ limits, the above discrete sum can be well approximated by replacing the summation by integration 
\begin{eqnarray}
Z_{G}(\b,\mu)\simeq\int dk~dN~\Omega(k,N)\exp-\b[E(k,N) -
\mu N]\label{zgapp}
\end{eqnarray}
Eq.(\ref{zgapp}) is the expression for the canonical partition function for a QIH
which is suitable for studying the thermal fluctuations. This may be regarded as the `thermalized' form of the quantum mechanical grand canonical partition function given by eq.(\ref{zgqm}). The information about the many-body quantum structure of the QIH is glossed by the classical behaviour of the the macroscopic variables.

\section{The Chemical Potential : IH thermodynamics from QIH thermodynamics}
\label{sec5}
It is worth mentioning that the most crucial consequence of all the above analysis, from the thermodynamic perspective, is that  the grand canonical partition function of QIH with $\mu=0$, is the canonical partition function for IH i.e. $Z_G^{QIH}(\b,0)=Z_C^{IH}(\b)$. This is nothing surprising because looking at the structure of the full Hilbert space of an IH, one would naturally consider the number of punctures to be unconserved for a given CS level$(k)$. Thus, in the quantum mechanical canonical partition function of an IH, the sum over all possible quantum states underlying a classical IH implies sum over {\it all possible} spin sequences which give rise to area eigenvalues within $\mathcal{O}(\lp^2)$ of the classical area irrespective of the number of punctures. In fact the partition function written in \cite{qg2}, although in the area ensemble, takes a sum over number of punctures also. But the significance of QIH and quantum hair $N$ remained unnoticed in\cite{qg2}. In that sense it is more reasonable to consider $N$ as a macroscopic variable and introduce $\mu$ on the first hand and then shift to the asymptotic view where the effect of the quantum hair disappears and $\mu$ is zero.

However, one must {\it not} confuse the analysis and the arguments regarding the chemical potential$(\mu)$ of a QIH with that of black holes having $\Phi \neq 0$ corresponding to any classical non-gravitational charge e.g. Reissner-Nordstrom, etc. One should remember that in this case there is an interplay between gravitational and non-gravitational fields to attain the equilibrium. Hence, it is obvious that $\Phi$ need not vanish to attain the equilibrium from the asymptotic viewpoint. The most important point to be noted is that one can tune the parameter $\Phi$ from outside by controlling the non-gravitational field, whereas, $\mu$ can not be controlled externally. The parameter $\Phi$ actually couples a nongravitational variable $(Q)$ to the partition function but $\mu$ does so in the case of $N$ which is purely (quantum) gravitational macroscopic variable. Although, $\Phi$ and $\mu$ appear to be playing the same role in different cases, basically they are very different as far as their physical implications are concerned. Further, $\Phi$ has nothing to do with quantum gravity, whereas, $\mu$ comes into play only when we consider the quantum gravity effects because the corresponding `charge' $N$ is a {\it quantum hair}\cite{GP,ampm2}. 

Finally, let us write down the IH partition function explicitly, which can be used directly to study the effects of Gaussian thermal fluctuations on the stability of a black hole\cite{th1}.  The chemical potential $\mu$ is defined as 
\ba
\mu=-\b^{-1}\p S_{MC}/\p N=-\s/\b
\ea 
Thus, from the asymptotic viewpoint, as discussed above, $\mu$ vanishes and consequently $\s=0$. Now, using $\s=0$ in eq.(\ref{siglm}), one obtains
\ba
1&=&\f{2}{\lm^2}\left(1+\f{\sqrt 3}{2}\lm\right)e^{-\f{\sqrt 3}{~2}\lm}~~~~\implies~~~~\lm=\lm_0
\ea
The above solution has been obtained graphically in \cite{meih} which gives the estimate $\lm_0=1.2$~.  Consequently, the microcanonical entropy is then given by 
\ba
S_{MC}&=&\lm_0~ \f{k}{2}\nn\\
&=&\lm_0~\f{A_{cl}}{8\pi\g\lp^2}~~~~~\text{using $k=A_{cl}/4\pi\g\lp^2$}\nn\\
&=&\f{A_{cl}}{4\lp^2}~~~~~~~~~\text{by choosing $\g=\lm_0/2\pi$}
\ea
One may note that, now we have $k=A_{cl}/2\lm_0\lp^2$. Also, as a parallel consequence, using $\lm=\lm_0$ in eq.(\ref{lmro}), one obtains
\ba
\f{k}{N}&=&1+\f{2}{\lm_0}+\f{4}{\lm_0(\sqrt 3\lm_0+2)}=c_0~\text{(say)}\label{knrel}
\ea
So, to calculate $Z_{IH}(\b)=Z^{QIH}_{G}(\b,0)$, it will not suffice to put $\mu=0$ in eq.(\ref{zgapp}), but we also have to impose the relation between $k$ and $N$ in eq.(\ref{knrel}) by using a Dirac delta function viz. $\d(k/c_0,N)$. Thus, eq.(\ref{zgapp}) now gets modified as
\ba
Z^{QIH}_{G}(\b,0)&\simeq&\int dk~dN~\Omega(k,N)\exp-\b E(k,N)~\d(k/c_0,N)\nn\\
&=&\f{1}{c_0} \int dk~\Omega(k)\exp-\b E(k)\nn\\
&=&\f{1}{2c_0\lm_0} \int dA_{cl}~\Omega(A_{cl})\exp-\b E(A_{cl})\label{thermcano}
\ea
where we have set $\lp^2=1$ for convenience. Thus, we obtain the canonical partition function for the IH i.e. $Z_{C}^{IH}(\b)$ beginning from the fundamental quantum structures of the QIH. This is the same partition function which was studied in \cite{th1}, and generalised to charged horizon in \cite{th2}, to study the effect of Gaussian thermal fluctuations on the stability of black holes. 
But, as has been pointed out earlier, the derivations of the partition functions were based on some approximations and failed to establish a clear and direct link with the undelying quantum theory. One can see Appendix \ref{sec6} for an elaborate discussion in this issue.

\section{Discussion}
\label{sec7}
The prime motive of the paper was the exact derivation of the horizon partition function using the fundamental quantum theory of the horizon geometry, which, we hope, has been elaborately worked out here up to a satisfactory level. The fact that the entropy of spacetime geometry vanishes in the absence of a horizon \cite{hawgib}, strongly motivates one to think of independent quantum degrees of freedom of the horizon giving rise to the thermodynamic properties uniquely associated with the horizon; and the derivation of the thermodynamic partition function right from the scratch, that has been presented in this work, satifactorily quenches the urge by establishing the all needed direct link between the quantum and thermodynamic aspects of the horizon. Hence, the major acquisition one can have by studying this work is the logical step by step derivation of the theromodynamic partition functions of the black hole horizon based on the underlying quantum geometric framework and a clear understanding about the role of quantum and thermal fluctuations in the corresponding thermodynamics. Following this, the `thermalized' form of the partition function derived here, which was previously used in the literature to study the thermal fluctuations of the black hole horizons, now has a clear meaning and stands on the firm ground provided by the quantum geometric description of black hole horizon\cite{qg1,qg2}. This can be considered as an important value addition to the literature of black hole horizon thermodynamics. Apart from this, an important clarification made in this paper is that the QIH thermodynamics is delicately different from IH thermodynamics due to the role of the quantum variable $N$ which is present in the quantum scenario, but absent in the classical thermodynamic picture. In fact, by beginning from IH framework one completely misses out the essence of the underlying quantum structure, resulting in the incorporation of ad hoc assumptions and unnecessary approximations in the procedure. As has been pointed out earlier, the identification of $N$ as a macroscopic variable for QIH is the difference. Now, looking at the full Hilbert space of an IH which allows arbitrary number of punctures for a given $k$, one may be tempted to ask that why at all shall we consider  $N$ as a macroscopic variable a priori. The answer to this question may be that the Hamiltonian operator\cite{me6,me7} commutes with the number operator for punctures which renders $N$ to be a constant of motion or, in other words, apparently there is no dynamics that that can change the number of punctures\footnote{This particular point is due to Daniele Oriti who placed this argument in favour of considering $N$ as a macroscopic variable for a QIH during a seminar at Albert Einstein Institute, Potsdam-Golm, Germany. This point was also made by Sumati Surya of Raman Research Institute, Bangalore.}. On the other hand, there is no role of $N$ at the classical level which is manifested by the absence of any $\mu\d N$ term in the first law derived from the classical theory of IH\cite{ih3}. Hence, it is most appropriate to begin from the study of quantum mechanical ensembles of QIHs by considering $N$ as a macroscopic variable and take the limit $\mu=0$ to arrive at the IH scenario, as shown in this paper. However, all the arguments regarding the role of $\mu$ in the local and asymptotic views remain at the qualitative level, the only understanding being the redshift factor as the cause of the behaviour of $\mu$. A quantitative understanding of how $\mu$ vanishes at the asymptopia, by an LQG calculation, may bring forth some new insights. Finally, it may be pointed out that, although we restrict our analysis to purely gravitational case and stop at canonical partition function of the IH, one can trivially consider non-gravitational charge and extend the analysis to the grand canonical partition function of charged IHs. One can look into \cite{th2} for this extension to charged horizons. But, one must be alert that there will be tricky issues involved if one tries generalize the analysis to rotating horizons, because we still do not have a proper quantum theory of rotating IH and this may be considered as a challenging research problem with potential consequences.

\vspace{0.5cm}
{\bf Acknowledgment :} I acknowledge the financial support provided by the Department of Atomic Energy of India during the initial stages of this work. The later stages of this work is funded my the Max-Planck Partner Group on Cosmology and Gravity. I sincerely thank an anonymous referee for giving valuable suggestions which has resulted in the improvement of the presentation of the manuscript.

\appendix
\section{Some physical speculations from the chemical potential }
The chemical potential for QIH is purely of quantum origin and is absent in the classical theory. This makes the role of the chemical potential quite subtle to understand and of course very different from our usual notion of chemical potential occurring in thermodynamics of ordinary systems.  It allows us to make some interesting speculations regarding the possible role of the chemical potential $\mu$ as far as the physical issues regarding the QIH are concerned and the possible new physics which may follow, in the following sections.

\subsection{Quantum inter-conversion of geometry and matter}
\label{subsec5.1}
In case of a system of gas of particles, we have an easy understanding of the physical meaning to the chemical potential $\mu$. It governs the physical exchange of particles between the system(the gas) and the ambient reservoir, both of which are 3+1 dimensional systems. Hence, most crucially, the particles retain their identity whether in the system or the reservoir and the scenario can be well visualized. But the scenario of QIHs is very much different. First of all the QIH is the quantum structure of the 2+1 dimensional IH, whereas the heat bath i.e. the exterior spacetime is 3+1 dimensional. The punctures, which are topological defects on the QIH are describable only as long as they are on the QIH. They are not defined elsewhere off the QIH. Thus there is no possible physical reservoir of punctures with which the QIH can physically {\it exchange} punctures. In this case the reservoir in contact with the QIH carries an abstract sense. Well, it can be argued that the punctures get detached to yield open ends in the bulk quantum geometry which are often considered as matter excitations\cite{lqgm}, but still they are not punctures which are quantum geometric excitations. The physical process in either direction is a conversion between two {\it different} objects -- i) detachment of punctures : conversion of horizon geometry to bulk matter ii) attachment of punctures : conversion of bulk matter to horizon geometry. Hence, the scenario of exchange of similar objects does not appear here. The very concept of `exchange' gets replaced by the concept of `conversion'.  To be more precise and to make comparison(or rather distinction) between the two apparently similar scenarios we can say that the fluctuations of $N$ imply `exchange of particles' for a gas and `mutual conversion of bulk matter and horizon geometry at the quantum level' in case of a QIH. It is quite clear that in case of QIHs the physical meaning of the parameter $\mu$ is drastically different from that in the case of a gas of particles. The situation urges us to understand the precise quantum dynamics which converts a geometrical excitation on the horizon into an excitation of matter field in the bulk.

\subsection{Quantum Topology of the Isolated Horizon : Local vs Asymptotic View}
\label{subsec5.2}
What we have been dealing with in this work is pure gravity and there is no non-gravitational field in this scenario. In the LQG framework, the degrees of freedom of the QIH are completely described by CS theory coupled to the topological defects on the horizon which act as sources (topological charges) \cite{qg1,qg2}. The macroscopic variables $k$ and $N$ are well defined as far as LQG is concerned and are pure gravity variables. Both the dynamical and the equilibrium states of the horizon are completely governed by purely quantum gravitational effects, devoid of any non-gravitational fields in this scenario. Thus, as opposed to the `tunable' potential corresponding to any non-gravitational charges, the `chemical potential' of a QIH corresponding to the topological charge $N$ can not be tuned externally. What we can do is to observe the behaviour of the parameter and understand its consequences as far as the thermodynamics of a QIH is concerned. Hence, even though we define the grand canonical partition function for a QIH by fixing $\b$ and $\mu$, but it may happen that a particular value $\mu$ is allowed for the equilibrium.

At thermodynamic equilibrium, the macroscopic variables of the QIH  are related by an equation of state. No matter crosses the horizon. But, even in this thermodynamic equilibrium condition a non-zero `chemical potential' i.e. $\mu\neq 0$ will imply a tendency of fluctuation of $N$ about the mean value $\Q$ is still dynamically present there.  Since the horizon has attained equilibrium i.e. isolated, the area is now fixed, the fluctuations of $N$ which can now take place are only those which can keep the area fixed. A bit of thinking will allow one to realize that these fluctuations of $N$ are nothing but the changes in the quantum topology of the QIH (e.g. {\it one} puncture of area $a$ get replaced by {\it two} punctures of area $a/2$), hence can be appropriately called quantum topological fluctuations. Thus, in spite of the fixed area at equilibrium, there will be a puncture dynamics going on for a nonzero chemical potential which gives rise to the visualization of the quantum description of radiation and accretion from the horizon\cite{kras}.

However, tracing back to the original proposal of the existence of quantum hair $N$ in \cite{GP}, it should be remembered that these quantum topological fluctuations will only be observable for a local observer close to the horizon and the effects must vanish for an asymptotic observer who can only see the classical IH, the existence of punctures and corresponding dynamics being glossed out at infinity. This explanation is supported by the fact that the first law of IH thermodynamics\cite{ih3} does not contain any $\mu \d N$ term as opposed to that of a QIH\cite{GP}. Thus what is observable at asymptopia is quantum topological equilibrium i.e. the smoothness of the topology of the horizon  $S^2\times R$. Such a quantum topological equilibrium will be attained only for $\mu=0$.

But for a local observer very close to the horizon, the quantum topological fluctuations must be observable and in that case a positive chemical potential\cite{meih} must be there which will result in a quantum topological fluctuation going on for a fixed classical area of the IH. This is like quantum dynamics underlying a classical equilibrium. The situation gives rise to the scenario that an infalling observer just before hitting the horizon will see a region of quantum fuzziness, which must be present there due to the quantum uncertainties evoked by the quantum structure of the IH. However, this explanation is provided on intuitive grounds and remain very much speculative unless one explores the scenario with LQG dynamics.  

\section{Value Addition to Earlier Literature -- Bridging the gap}
\label{sec6}
Here we discuss the different background independent statistical mechanical approaches to black hole horizon thermodynamics that has gradually evolved step by step in literature, which will finally give us an understanding about the shortcomings of these earlier approaches and how the related technical caveats get eliminated by the present work based on the QIH framework. This will reveal the value addition of the present work to the existing literature. 

As far as our knowledge is concerned, a generic statistical mechanical approach without prior use of any background metric was used to study the canonical ensemble scenario with the aim to investigate the effects of thermal fluctuations for black holes in \cite{dmb,cm4,gomed}. The canonical partition function was written as
\ba
Z_C&=&\sum_ig(E_i)\exp-\b E_i\simeq \int g(E)\exp -\b E~dE\nn
\ea
 for a black hole(assuming in the hindsight). This is tantamount to {\it model} the quantum theory of the black hole as that of a single particle, alike the harmonic oscillator, whose energy spectrum is characterized by a single quantum number and also equispaced. Without the equispaced energy spectrum, the discrete sum can not be approximated as an integration in any limit.  The usual thermodynamic results following from the effective single particle model were applied to selected black holes and relevant conclusions were drawn. The above partition function, if not stated to be only a heuristic {\it model}, has hardly any relationship with actual black hole quantum states originating from the theory of QIH.
The above {\it model} got improved a bit and some properties of QIH were incorporated in \cite{cm1}. The canonical partition function was written as 
\ba
Z_C&=&\sum_ig[E(A_i)]\exp-\b E(A_i)\simeq\int dx~g[E(A(x))]\exp-\b E[A(x)]\nn
\ea
considering that the energy is a function of the area of the horizon and then, assuming that all the punctures on the horizon carry spin-half (neglecting the effects of other spins), the area spectrum of the horizon is made equispaced i.e. $A\propto N$. The energy states are now characterized by a single quantum number but the energy spectrum is no more equispaced. Thus, the effective single particle model for black holes has got more generalized and for large horizons with sufficiently densely packed quantum states the discrete sum over the quantum number is replaced by an integration over a continuous variable which represents the quantum number in the continuum limit. It was only in \cite{th1} that a logical explanation of the partition function was given starting from the IH framework, albeit in a superficial approach using the tools of LQG. The partition function was no more any model but a true partition function derived from the fundamental theory of IH. But still the technical assumptions and approximations regarding the linearized area spectrum, etc. plagued the calculations resulting in the effective one particle picture.

Similar to the canonical ensemble study of the effective one-particle model, the fluctuations of {\it non-gravitational} charges were studied in the grand canonical ensemble  in \cite{cm2,gomed}, considering that the effective one-particle carries a quantized charge representing the non-gravitational charge of the black hole. The partition function is written as
\ba
Z_G&=&\sum_ig[E(A_i), Q_j]\exp-\b E(A_i,Q_j)+\b \phi Q_j\nn\\
&\simeq&\int dx~dy~g[E(A(x)), Q(y)]\exp-\b E(A(x),Q(y))+\b \phi Q(y)\nn
\ea
using similar assumptions and approximations in the appropriate limits as discussed above for the canonical case. In \cite{th2} the partition function was derived in the IH framework, albeit in a superficial approach using the tools of LQG. The explanation of the Hamiltonian associated with the horizon was improved and the structure of the formalism was put on a stronger ground than ever. The partition function was a generic one for charged IH. But still the technical assumptions and approximations regarding the linearized area spectrum, etc. plagued the calculations resulting in the effective one particle picture. It is worth explaining that the assumption of the linearized quantum area spectrum is tantamount to disregard the role of $N$ as an independent macroscopic parameter for QIH. Hence, the existence of the macroscopic variable at the quantum level continued to be opaqued by the assumption of the linearized quantum area spectrum.

The relevance of the many-particle nature of the QIH remained unnoticed and hence, the canonical ensemble of QIHs and IHs have not been differentiated. It was in \cite{GP} that the crucial observation was made and was justified with simple reasonings in \cite{ampm2}, that the number of punctures on the QIH should be considered to be an independent macroscopic variable which characterizes a macrostate of a QIH. Hence, it is worth investigating the crucial difference between the canonical ensemble thermodynamics of QIH and classical IH. The canonical partition function is defined for a fixed $N$ for QIH, whereas for classical IH there is no concept of $N$. That is why the canonical partition function in eq.(\ref{zcapp}) is tagged by the parameter $N$, where as there is no such tag for the canonical partition function of IH in \cite{th1}. In case of IH the canonical partition function is defined for fixed non-gravitational charges\cite{th1} which are allowed to fluctuate in the corresponding grand canonical ensemble\cite{th2}. But for the QIH, the many-body quantum structure compels us to treat the complete thermodynamics in the grand canonical ensemble even in the absence of non-gravitational charges.

Further there are some very important consequences of the study of the thermodynamics of the black holes beginning from the fundamental quantum structure of QIH rather than from the classical IH framework. For a canonical ensemble of IH\cite{th1} the energy is considered to be a function of area which is again assumed to be equispaced so that the summation can be replaced by integration. This assumption is severe and requires that only similar spins can be considered to be there at each puncture, preferably spin half, which is not true, although statistically spin half is the majority as found from the most probable distribution in the microcanonical ensemble. In the usual background independent approach to study the horizon thermodynamics in \cite{th1,th2,cm1,cm2,cm3} it is considered that `the energy is a function of area'. In a precise sense, this is a very confusing statement. The statement should be made very precise, otherwise the underlying essence is not captured. Suppose one says the classical energy of an IH is the function of its classical area(or equivalently the CS level), which is fine. But this is not any sort of energy spectrum, rather the mean energy from the perspective of the QIH scenario\cite{me6,me7}. Thus the statement `equispaced  area spectrum' does not have a clear meaning in the context of {\it classical} IH and hence approximations and assumptions had to be incorporated in \cite{th1,th2,cm1,cm2,cm3}.

In this work, as we begin from the very fundamental quantum structure of the IH i.e. the QIH framework, the physical and technical aspects of the calculations are logical and straightforward, devoid of any assumption and approximation. It is very much transparent that how the mean energy shows up in the partition function even though we begin from the quantum mechanical ensemble using the Hamiltonian operator for the QIH and hence the true energy spectrum of the same.  Moreover, without having to make any assumption, we have the mean energy as the function of $k$ and $N$, which are integers and hence equispaced. Thus the thermalized forms of the partition functions automatically reduce to the convenient forms which can further be approximated by replacing summation by integration.


\begin{thebibliography}{777}

\bibitem{hawgib} G. W. Gibbons and S. W. Hawking, {\it Phys. Rev.} {\bf D15}, 2752 (1977)


\bibitem{th1} P. Majumdar, {\it Class. Quant. Grav.} {\bf 24} (2007) 1747; \href{http://arxiv.org/abs/gr-qc/0701014}{arXiv:gr-qc/0701014}


\bibitem{th2} A. Majhi and P. Majumdar, {\it Class. Quant. Grav.} {\bf 29} (2012) 135013; \href{http://arxiv.org/abs/1108.4670}{	arXiv:1108.4670v1}


\bibitem{bhal} J. D. Bekenstein, {\it Phys. Rev.} {\bf D7} (1973) 2333–2346; S. W. Hawking, {\it Phys. Rev.} {\bf D13} (1976) 191–197  

\bibitem{hrad} S. W. Hawking, {\it Commun. Math. Phys.} {\bf 43}, 199--220 (1975)

\bibitem{wrad} R. M. Wald, {\it Commun. Math. Phys.} {\bf 45}, 9--34 (1975)

\bibitem{laws} J. M. Bardeen, B. Carter and S. W. Hawking, {\it Commun. math. Phys.} {\bf 31}, 161-170 (1973)


\bibitem{dmb} S. Das, P. Majumdar and R. K. Bhaduri, {\it Class. Quant. Grav.} {\bf 19} (2002) 2355-2368, \href{http://arxiv.org/abs/hep-th/0111001v3}{arXiv:hep-th/0111001}


\bibitem{gomed} G. Gour and A. J. M. Medved, {\it Class. Quant. Grav.} {\bf 20} (2003) 3307–3326, \href{http://arxiv.org/abs/gr-qc/0305018v2}{arXiv:gr-qc/0305018}



\bibitem{cm1} A. Chatterjee and P. Majumdar, {\it Phys. Rev. Lett.} {\bf 92} (2004) 141031; \href{http://arxiv.org/abs/gr-qc/0309026}{arXiv: gr-qc/0309026}. 


\bibitem{cm2}  A. Chatterjee and P. Majumdar, {\it Phys. Rev.} {\bf D71} (2005) 024003; \href{http://arxiv.org/abs/gr-qc/0409097}{arXiv: gr-qc/0409097}.


\bibitem{cm3} A. Chatterjee and P. Majumdar, {\it Phys. Rev.} {\bf D72} (2005) 044005; \href{http://arxiv.org/abs/gr-qc/0504064}{arXiv: gr-qc/0504064}.


\bibitem{cm4} A. Chatterjee and P. Majumdar, {\it Black hole entropy: quantum vs thermal fluctuations}, \href{http://arxiv.org/abs/gr-qc/0303030v1}{arXiv:gr-qc/0303030}


\bibitem{qg1} A. Ashtekar, J. Baez, A. Corichi and K. Krasnov, {\it Phys. Rev. Lett.} {\bf 80} (1998) 904-907; \href{http://arxiv.org/abs/gr-qc/9710007v1}{arXiv:gr-qc/9710007v1}


\bibitem{qg2} A. Ashtekar, J. Baez and K. Krasnov, {\it Adv. Theor. Math. Phys.} {\bf 4} (2000) 1; \href{http://arxiv.org/abs/gr-qc/0005126}{arXiv:gr-qc/0005126v1}



\bibitem{ih1} A. Ashtekar, A. Corichi and K. Krasnov, {\it Adv. Theor. Math. Phys.} {\bf 3} (2000) 419-478; \href{http://arxiv.org/abs/gr-qc/9905089v3}{arXiv:gr-qc/9905089v3}


\bibitem{ih2} A. Ashtekar, C. Beetle and S. Fairhurst, {\it Class. Quant. Grav.} {\bf 17} (2000) 253-298; \href{http://arxiv.org/abs/gr-qc/9907068}{arXiv:gr-qc/9907068v2}


\bibitem{ih3} A. Ashtekar, S. Fairhurst and B. Krishnan, {\it Phys. Rev.} {\bf D62} (2000) 104025; \href{http://arxiv.org/abs/gr-qc/0005083}{arXiv:gr-qc/0005083v3}


\bibitem{ih4} A. Ashtekar and B. Krishnan, {\it Living Rev. Rel.} {\bf 7} (2004) 10; \href{http://arxiv.org/abs/gr-qc/0407042}{arXiv:gr-qc/0407042v3}


\bibitem{ih5}A. Ashtekar, C. Beetle, O. Dreyer, S. Fairhurst, B. Krishnan, J. Lewandowski and J. Wisniewski, {\it Phys. Rev. Lett.} {\bf 85} (2000) 3564-3567, \href{http://arxiv.org/abs/gr-qc/0006006}{arXiv:gr-qc/0006006v2}

\bibitem{km11} R. K. Kaul and P. Majumdar, {\it Phys. Rev.} {\bf D83} (2011) 024038; \href{http://arxiv.org/abs/1004.5487}{arXiv:1004.5487}



\bibitem{ashlew} A. Ashtekar and J. Lewandowski, {\it Class. Quantum Grav.} {\bf 21} (2004) R53, \href{http://arxiv.org/abs/gr-qc/0404018v2}{arXiv:gr-qc/0404018v2}



\bibitem{meih} A. Majhi,  {\it Class. Quant. Grav.} 31 (2014) 095002, \href{http://arxiv.org/abs/1205.3487}{arXiv:1205.3487}


\bibitem{GP} A.Ghosh and A.Perez, {\it Phys. Rev. Lett.} {\bf 107}, 241301 (2011); \href{http://arxiv.org/abs/1107.1320}{arXiv:1107.1320v2}  


\bibitem{ampm2} A. Majhi and P. Majumdar, {\it Quantum Hairs and Isolated Horizon Entropy from Chern-Simons Theory}, {\it Class. Quant. Grav.}, (to be published), \href{http://arxiv.org/abs/1301.4553}{arXiv:1301.4553}


\bibitem{km00} R. K. Kaul and P. Majumdar, {\it Phys. Rev. Lett.} {\bf 84} (2000) 5255-5257; \href{http://arxiv.org/abs/gr-qc/0002040}{arXiv:gr-qc/0002040v3} 

\bibitem{SU(2)} The fact that the CS theory on the IH is in fact SU(2) gauge invariant rather than U(1) was first pointed out in \cite{km00}...


\bibitem{kras}K. V. Krasnov, {\it Gen. Rel. Grav.} {\bf 30} (1998) 53-68; \href{http://arxiv.org/abs/gr-qc/9605047}{arXiv:gr-qc/9605047v3}; D. Pranzetti, {\it Phys. Rev. Lett.} {\bf 109}, (2012) 011301 , \href{http://arxiv.org/abs/1204.0702}{arXiv:1204.0702v1}; {\it  Class. Quant. Grav.} {\bf 30} 165004 (2013), \href{http://arxiv.org/abs/1211.2702}{arXiv:1211.2702v1} 



\bibitem{landau} L. D. Landau and E. M. Lifschitz, {\it Statistical Physics}, Pergamon Press, 1980



\bibitem{me6} A. Majhi, {\it Phys. Rev.} {\bf D88} 024010 (2013), \href{http://arxiv.org/abs/1303.4829}{arXiv:1303.4829}


\bibitem{me7} A. Majhi, {\it Quantum to Clasical Isolated Horizon : Energy Spectrum of equilibrium Black Holes},  \href{http://arxiv.org/abs/1303.4832}{arXiv:1303.4832}




\bibitem{lqgm} Hugo A. Morales-Tecotl and Carlo Rovelli, {\it Phys. Rev. Lett.} {\bf 72} (1994) 3642-3645, \href{http://arxiv.org/abs/gr-qc/9401011}{arXiv:gr-qc/9401011}

\bibitem{km98} R. K. Kaul and P. Majumdar, {\it Phys.Lett.} {\bf B439} (1998) 267-270; \href{http://arxiv.org/abs/gr-qc/9801080}{arXiv:gr-qc/9801080v2}



\end{thebibliography}
\end{document}